\documentstyle[aps,epsfig,prl,floats]{revtex}

\tightenlines
\begin{document}
\draft

\title{Spectral weight
contributions of many-particle bound states and continuum}

\author{
         Weihong Zheng$^{(a)}$\cite{zwh},
        Chris J.~Hamer$^{(a)}$\cite{cjh}, and
        Rajiv R.~P.~Singh$^{(b)}$
}
\address{
$^{(a)}$ School of Physics, University of New South Wales, Sydney NSW
2052, Australia\\
$^{(b)}$ Department of Physics, University of California, Davis, CA
95616\\
}

%\author{Weihong Zheng}
%\email[]{w.zheng@unsw.edu.au}
%\homepage[]{http://www.phys.unsw.edu.au/~zwh}
%\affiliation{School of Physics,
%The University of New South Wales,
%Sydney, NSW 2052, Australia.}
%
%\author{C.J. Hamer}
%\email[]{c.hamer@unsw.edu.au}
%\affiliation{School of Physics,
%The University of New South Wales,
%Sydney, NSW 2052, Australia.}
%
%\author{R.R.P. Singh}
%\email[]{singh@raman.ucdavis.edu}
%\affiliation{Department of Physics, University of California, Davis,
%CA95616, USA}

\twocolumn[\hsize\textwidth\columnwidth\hsize\csname
@twocolumnfalse\endcsname

\date{\today}
\maketitle

\begin{abstract}
Cluster expansion methods are developed for calculating the
spectral weight contributions of multiparticle excitations - continuum and
bound states - to high orders. A complete 11th order calculation
is carried out for the alternating Heisenberg chain. 
For $\lambda=0.27$, relevant
to the material $Cu(NO_3)_2.2.5D_2O$, we present detailed spectral weights
for the two-triplet
continuum and all bound states. We also examine variation
of the relative weights of one and two-particle states
with bond alternation from the dimerized to the uniform chain limit.
\end{abstract}

%\keywords{}

]

\narrowtext
%\section{Introduction}

In recent years there has been a growing interest in understanding
quantitatively the single and multiparticle
excitation spectra in quantum spin systems.
%From a theoretical point of view,
On the one hand, this understanding is important
in assessing the extent of a continuum due to conventional many-particle
excitations as compared to more exotic scenarios \cite{fraction}.
%From an experimental point of view,
On the other hand,
% the synthesis of many low-dimensional spin-half antiferromagnets and
the increased frequency and wave-vector
resolution of inelastic neutron scattering experiments means that
such spectra are being observed in real materials 
and need quantitative theoretical support \cite{ten02}.

Controlled and systematic calculation of the spin dynamics of quantum
spin models remains a challenging computational task.
Despite much progress in developing computational methods
%such as density matrix
%renormalization group and quantum Monte Carlo,
the multiparticle excitations remain poorly understood.

Here, we develop a general linked-cluster formalism to calculate
the single-particle and multi-particle contributions to
the dynamical structure factor by means of high-order series expansions.
 We apply the method to the
alternating Heisenberg chain (AHC) model, where expansions are made around
the strong coupling limit of decoupled spin dimers.

We follow the formalism of Tennant et al.\cite{ten02}. The inelastic
neutron scattering cross-section\cite{mar71} at temperature $T = 0$
is proportional to the neutron scattering ``structure factor" $S^{\alpha\beta}({\bf
k},\omega)$
\begin{eqnarray}
S^{\alpha\beta}({\bf k},\omega) & = & \frac{1}{2\pi N} \sum_{i,j}\int
dt\exp[i(\omega t + {\bf k}\cdot ({\bf r}_i - {\bf r}_j)] \nonumber \\
 &  & \langle \psi_0 | S^{\alpha}_j (t)
S^{\beta}_i (0) | \psi_0 \rangle
\label{eq2a}
\end{eqnarray}
where ${\bf k}$ is the wavevector transfer, $\omega$ is the energy,
N is the number of scattering centres,
$\alpha,\beta = x,y,z$ are Cartesian coordinates, $i,j$ label sites
of the system,
and $| \psi_0 \rangle $ is the ground state of the Hamiltonian.
This is just
the space and time Fourier transform of the
spin-spin correlation function.
For AHC,
the structure factor has only one independent spin component, and
henceforth
we concentrate our attention on $S^{-+}({\bf k},\omega)$.

Integrating equation (\ref{eq2a}) over energy, we get
the integrated structure factor, i. e.
the Fourier transform of the spin-spin correlation
function at $t = 0$:
\begin{equation}
S^{-+}({\bf k}) = \sum_{{\bf r}} C^{-+}({\bf r})\exp[i{\bf k\cdot r}]
\label{eq7}
\end{equation}

Inserting a complete set of momentum eigenstates 
${ | \psi_n %({\bf k}) 
\rangle }$ of $H$
between the spin operators in equation (\ref{eq2a}), and using
translation invariance, we can express the
spin structure factor as a sum over ``exclusive" structure factors

\begin{equation}
S^{-+}({\bf k},\omega) = \sum_{n} S_{n}^{-+}({\bf k},\omega),
\label{eq8}
\end{equation}
\begin{eqnarray}
S^{-+}_{n}({\bf k},\omega) & = & \frac{N_c^2}{N} \delta (\omega
 - E_{n} + E_0) \nonumber \\
 & & {\Big |} \sum_{i*}
\langle \psi_{n} ({\bf k}) | S_{i*}^{+} (0) | \psi_0 \rangle
 \exp[i {\bf k}\cdot {\bf r}_{i*}] {\Big |}^2
\label{eq12}
\end{eqnarray}
where the sum runs over all sites
$i^{*}$ in the unit cell, and
$N_c$ is the number of unit cells on the lattice. Barnes et
al.\cite{bar99} define the ``reduced exclusive structure factor" as

\begin{equation}
S^{-+}_{n}({\bf k}) = N_c
 {\Big |} \sum_{i*}
\langle \psi_{n} ({\bf k}) | S_{i*}^{+}  | \psi_0 \rangle
 \exp[i {\bf k}\cdot {\bf r}_{i*}] {\Big |}^2
\label{eq13}
\end{equation}
Each exclusive structure factor $ S_{n}^{\alpha\beta}({\bf q},
\omega)$ gives the intensity of scattering from $| \psi_0 \rangle$ to a
specific triplet excited state $\vert \psi_n ({\bf k}) \rangle$ \cite{bar99}.

We turn now to a discussion of algorithms for the calculation of
exclusive structure factors within perturbation theory.
Efficient linked cluster expansion methods have long been known\cite{gel00}
for calculating bulk properties of a quantum lattice system. Similar
methods for the calculation of 1-particle spectra were developed by
Gelfand\cite{gel96}, and were recently extended to 2-particle spectra by
Trebst et al\cite{tre00,zhe01}. Series calculations of exclusive 1-particle
structure factors or spectral weights appear in several places
(e.g. \cite{gel89,fle97,sin99});
and some low-order calculations for 2-particle states have recently been
made\cite{bar99,kne01,ten02}.
Knetter et al\cite{kne01} have used an
alternative approach based on `continuous unitary transformations' which
is also capable of giving bound state energy spectra and structure
factors to high order and in great detail.

Let us suppose that the Hamiltonian can be decomposed
\begin{equation}
H = H_0 + \lambda V
\end{equation}
where $H_0$ is the unperturbed Hamiltonian and $V$ is to be treated as a
perturbation. We aim to expand the multiparticle dispersion relations
and structure factors in powers of the parameter $\lambda$. For
illustrative purposes, we shall use the language of the AHC, but the
formalism can be applied more generally.

At zeroth order, the `single-particle' excitations consist
of triplets on a single dimer which can be labelled $ | m,
\alpha \rangle$, where $m$ labels the position of the excited dimer, and $
\alpha$ labels the angular momentum eigenstate (i.e. $S^2, S_z$).
The reduced exclusive structure factor is
\begin{equation}
S^{-+}(k) = {\Big |} \sum_{m,i^*,\alpha}
 \langle m,\alpha | S^{+}_{i^*} | 0 \rangle \exp[ik(x_{i^*} - x_m)]
 {\Big |}^2
\label{eq21}
\end{equation}

It is easy to see that the matrix element
$\langle m,\alpha | S^{+}_{i} | 0 \rangle $ $\equiv S^{+}_{1\alpha}(m,i)$
must obey a simple `cluster addition' property,
such that the elements $S^{+}_{1\alpha}(m,i)$ admit a linked cluster
expansion
\begin{equation}
S^{+}_{1\alpha}(m,i) = \sum_{\gamma \ni (m,i)}s^{\gamma,+}_{1\alpha}(m,i)
\label{eq23}
\end{equation}
where the sum over $\gamma$ denotes a sum over all connected clusters
which are ``rooted to", or contain, the positions of dimer $m$ and site
$i$. Correspondingly, the perturbation series expansion for
$S^{+}_{1\alpha}(m,i)$ could be formulated in terms of a diagrammatic
expansion where only connected diagrams contribute, although we will not
elaborate on this approach here.

An efficient linked cluster algorithm for calculation of the structure
factors can now be formulated, following Trebst et al.\cite{tre00,zhe01}:

i) Generate a list of connected clusters $\gamma$ appropriate to
the problem at hand (in the present case, they will simply consist of
chains of dimers of different lengths);

ii) For each cluster $\gamma$, construct matrices for the
Hamiltonian $H$ and spin operators $S^{+}_i$ in the basis of singlet and
triplet dimer states corresponding to $H_0$;

iii) `Block diagonalize' the Hamiltonian by an orthogonal
transformation
as outlined by Trebst et al\cite{tre00,zhe01}, constructed order-by-order in
perturbation theory so that the 1-particle states sit in a block by
themselves; simultaneously, transform the matrices for the spin
operators
\begin{equation}
 H^{{\rm eff}} = O^T H O, \hspace{2mm} S^{{\rm eff},+}_i = O^T S^{+}_i O
\label{eq25}
\end{equation}

iv) Subtract all sub-cluster contributions;

v) Insert in equation (\ref{eq23}), and then equation
(\ref{eq21}), to build up the exclusive structure factors order-by-order
in perturbation theory.

Since two triplets can combine to give total spin 1, there will also be
non-zero spin structure factors for 2-particle states. The 2-particle
states can be labelled according to their unperturbed counterparts in
position space $ | m,n;\alpha,\beta \rangle$ where $m,n$ label the two
dimer positions, and $\alpha,\beta$ the corresponding angular momentum
states.
Then the exclusive matrix element is
\begin{equation}
 \langle \psi_j | S^{+}_i | 0 \rangle =  \sum_{m,n}
f^{*j}_{\alpha\beta}(m,n) \langle m,n;\alpha,\beta | S^{+}_i | 0
\rangle
\label{eq28}
\end{equation}
where the 2-particle wavefunctions $f^j_{\alpha\beta}(m,n)$ can be
computed by the block diagonalization procedure of Trebst et
al.\cite{tre00,zhe01}.
The 2-particle matrix elements in position space can then be found by a
linked cluster algorithm essentially the same as that for the 1-particle
states.

We have used this method to investigate the spectral weights for
the  alternating Heisenberg chain Hamiltonian, described
by the following Hamiltonian
\begin{equation}
H =  \sum_{i} {\Big [} {\bf S}_{2i}\cdot {\bf S}_{2i+1} + \lambda {\bf S}_{2i-1}\cdot {\bf S}_{2i} 
{\Big ]} \label{eqH}
\end{equation}
where the ${\bf S}_i$ are  spin-$\frac{1}{2}$ operators at site $i$,
 and $\lambda$ is the alternating dimerization.

There is a considerable literature on this model, which has been reviewed
recently by Barnes et al.\cite{bar99}.
At $\lambda = 0$, the system consists of a chain of decoupled dimers,
and in the ground state each dimer is in a singlet
state, while excited states are made up from the three triplet excited
states on each dimer, with a finite energy gap.
At $\lambda = 1$, we regain the uniform Heisenberg chain,
which is gapless.
%Several theoretical papers\cite{den79,cro79,bla81,
%uhr99,aff97,sor98} have discussed the
%approach to the uniform limit and the way in which the alternating
%system with spin-1 magnon excitations crosses over to the uniform system
%with spin-1/2 `spinon' excitations. [Details?]
Numerical studies of the model include series expansions\cite{duf68,bon82,sin99}, and
exact diagonalizations for finite lattices\cite{soo85,spr86}.

\begin{figure}
\begin{center}
% \vskip -7mm
 \epsfig{file=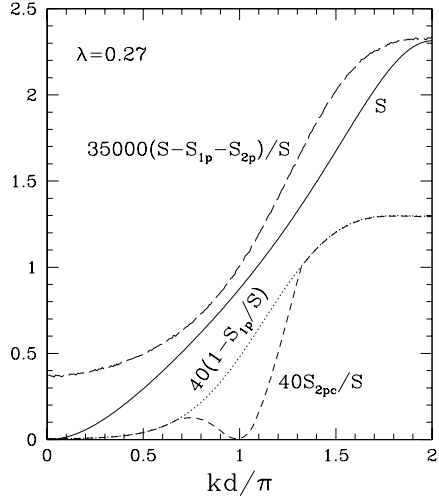,width=7cm}
  \vskip 5mm
 \caption[]
         {The integrated structure factor $S$, and some relative spectral weights
          at $\lambda=0.27$, as functions of wavevector k, where d is
the distance between dimers.}
 \label{fig_1}
 \end{center}
\end{figure}

The 2-magnon bound states were previously studied by Uhrig and
Schulz\cite{uhr96} using an RPA approach. They found a singlet bound state
below the 2-particle continuum for all momenta $k$ and over the whole
range of $\lambda < 1$. They also predicted a triplet bound state and a
quintuplet antibound state near $k = \pi/2$ for $\lambda$ not too large.
These conclusions were supported in later studies
\cite{bou98,fle97,she99}. Trebst et al.\cite{tre00,zhe01a} found
in a high-order series expansion study that in fact there are {\it two}
$S = 0$ ($S_1$ and $S_2$) and two $S = 1$ bound states ($T_1$ and $T_2$), 
together with two $S = 2$
antibound states ($Q_1$ and $Q_2$), if one goes sufficiently close to $k = \pi/2$.
Hence the model
displays some interesting multi-particle dynamics which can be explored
both theoretically and in experiments. Neutron scattering experiments,
however, will only be sensitive to the triplet bound states.

Barnes et al.\cite{bar99} and Tennant et al.\cite{ten02} have shown how to calculate
exclusive structure factors for the 2-particle states by low-order
series expansions, and have made a comparison with experimental data\cite{xu00,ten02} for
the copper nitrate material, $Cu(NO_3)_2.2.5D_2O$.

Here we extend the series expansions to high orders.
Series have been computed up to order $\lambda^{11}$
for the integrated structure factor $S(k)$,
and the exclusive structure factors for the 1-particle triplet state $S_{1p}(k)$,
the two 2-particle triplet bound states, $S_{T_1}$ and $S_{T_2}$,
the 2-particle
continuum $S_{2pc}$, and finally
the total 2-particle structure factor (the sum rule)
$S_{2p}$.
The 1-particle structure factor has been computed to
order $\lambda^3$ by Barnes {\it et al.}\cite{bar99}, but
our series disagree with their results from second order.

\begin{figure}
 \begin{center}
 \vskip -7mm
 \epsfig{file=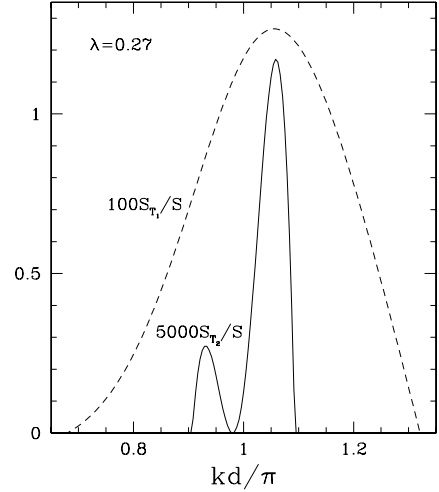, width=7cm}
  \vskip 5mm
 \caption[]
         {The relative spectral weight of the  2-particle triplet bound states
$T_1$ and $T_2$
          at $\lambda=0.27$.}
 \label{fig_2}
 \end{center}
\end{figure}

The integrated structure factor for $\lambda=0.27$, relevant
to the material $Cu(NO_3)_2.2.5D_2O$, is shown as a function of wavector
$k$ in
Fig. \ref{fig_1}, along with some relative spectral weights. It can be
seen that the structure factor peaks strongly at $kd = 2\pi$, and is
dominated by the 1-particle state, which carries at least
96\% of the total weight.
The relative weight for more than two particles (i.e. $(S-S_{1p}-S_{2p})/S$ ) is less than
0.0069\%.

The spectral weights for 2-particle triplet bound states $T_1$ and $T_2$
are given in Fig. \ref{fig_2}. We know from previous studies that bound states
exist only in a limited range of momenta near $kd=\pi$, and the spectral weight
reflects this fact.
The maximum relative spectral weight for $T_1$ is about 1.2\%
at $kd = \pi$,
while the spectral weight for $T_2$ has two peaks, with a zreo in
between them, and has a very small
maximum relative weight
about  0.023\%. The two-peaked structure can be traced back to the
bound-state wave function for $\lambda \to 0$, where the two triplets
are separated by an odd number of dimers\cite{zhe01a}.

\begin{figure}
 \begin{center}
 \vskip -7mm
 \epsfig{file=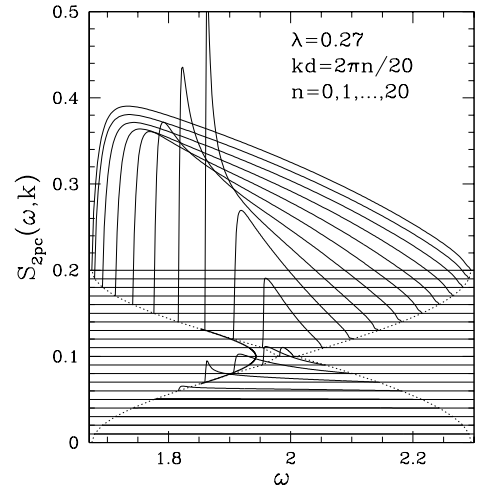, width=7cm}
  \vskip 5mm
 \caption[]
         {The structure factor (shifted by $n/100$) for the 2-particle continuum versus energy $\omega$
          at $\lambda=0.27$ and $kd=2\pi n/20$, $n=0,1,2,\cdots,20$.}
 \label{fig_3}
 \end{center}
\end{figure}

Figure \ref{fig_3} shows the structure factor for the 2-particle continuum,
also at $\lambda=0.27$, sliced at various momentum intervals. Also shown
as a solid curve is the dispersion relation for the triplet bound state
$T_1$. It can be seen that where the bound state merges with the
continuum, a very sharp peak develops at the lower edge of the
continuum.

\begin{figure}
 \begin{center}
 \vskip -7mm
 \epsfig{file=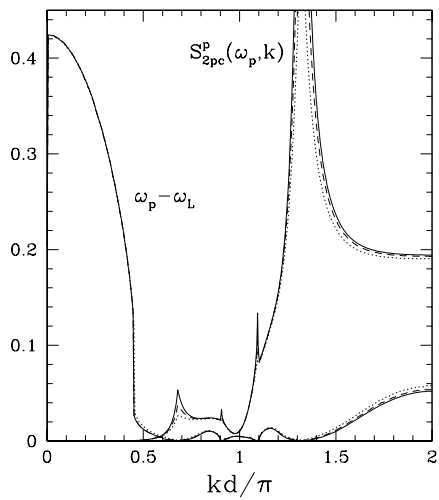, width=7cm}
  \vskip 5mm
 \caption[]
{The peak value of the 2-particle continuum structure factor $S_{2pc}^{\rm p}(\omega_p, k)$, and
         the difference ($\omega_p - \omega_L$) between the position of
the peak
 and the lower edge of the continuum,
graphed as functions of $k$
for finite lattices of 600 (dotted line), 1200 (dashed line) and
2400 (solid line) sites
          at $\lambda=0.27$.}
\label{fig_4}
\end{center}
\end{figure}

Figure \ref{fig_4} shows first, the {\it position} of the peak in the
continuum structure factor relative to the lower edge, $\omega_p
-\omega_L$, and secondly the peak {\it value} of the structure factor
$S_{2pc}^{\rm p}(\omega_p, k)$, as functions of $k$, calculated with
finite lattices of 600, 1200 and 2400 sites in the Fourier transform
\cite{zhe01}).
It can be seen that at the threshold points where $T_1$ and $T_2$ merge
with the continuum,
($\omega_p-\omega_L$) goes quadratically to zero, and the peak value
$S_{2pc}^{\rm p}(\omega_p, k)$ shows spikes which appear to be diverging
to infinity for the bulk system.

\begin{figure}
 \begin{center}
 \vskip -7mm
 \epsfig{file=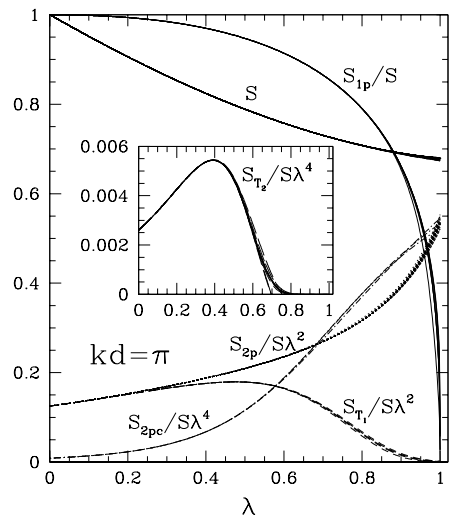, width=7cm}
  \vskip 5mm
 \caption[]
         {The relative spectral weights for 1-particle $S_{1p}$, 
         all 2-particle states $S_{2p}$, 
individual
         bound states $T_1$ and $T_2$, and the continuum, at $kd=\pi$.}
\label{fig_5}
\end{center}
\end{figure}

We have also studied the behaviour of the spectral weight as a function of
$\lambda$.
%Figure \ref{fig_1p_lambda} shows the relative spectral weight of
%the 1-particle triplet state, which
%approaches zero as $\lambda\to 1$, as expected.
%The total 2-particle spectral weight, and those for the individual
%2-particle triplet bound states $T_1$ and $T_2$ and continuum
%for $kd=\pi$
%are given in Figure \ref{fig_5}. It can be seen that the spectral
%weight for
%$T_2$ vanishes before $\lambda=1$,
%while that for $T_1$ vanishes at $\lambda=1$.
%The 2-particle
%continuum has about 54\% weight at $\lambda = 1$, implying a large
%overlap between the two-triplet and the two-spinon descriptions\cite{karbach}
%in the uniform limit.
Figure \ref{fig_5} shows, for $kd=\pi$, the integrated structure 
factor $S$,
the relative spectral weight of
the 1-particle triplet state, 
the total 2-particle spectral weight, and those for the individual
2-particle triplet bound states $T_1$ and $T_2$ and continuum.
 It can be seen that 
the relative spectral weight of
the 1-particle triplet state
approaches zero as $\lambda\to 1$, as expected, and
the spectral weight for
$T_2$ vanishes before $\lambda=1$,
while that for $T_1$ vanishes at $\lambda=1$.
The 2-particle
continuum has about 54\% weight at $\lambda = 1$, implying a large
overlap between the two-triplet and the two-spinon descriptions\cite{karbach}
in the uniform limit.

%\begin{figure}
% \begin{center}
% \vskip -7mm
% \epsfig{file=Weight_x_S1p_o_S_4_sel_k.ps, width=7cm}
%  \vskip 5mm
% \caption[]
%         {The relative spectral weight for the 1-particle triplet state at $kd=\pi/2, \pi, 3\pi/2, 2\pi$.}
%\label{fig_1p_lambda}
%\end{center}
%\end{figure}

These results show that detailed and accurate results for structure
factors and spectral weights can be obtained using high-order series
expansions, as was also demonstrated for the spin-ladder
model by Knetter et al.\cite{kne01}.

This work is supported by a grant
from the Australian Research Council and by US National Science Foundation
grant number DMR-9986948. We have benefited from discussions with
Professor O.P. Sushkov and Professor J. Oitmaa.
The computation has been performed on an AlphaServer SC
 computer. We are grateful for the computing resources provided
 by the Australian Partnership for Advanced Computing (APAC)
National Facility.

\bibliography{basename of .bib file}

% \newpage

%=======================================================================
\end{document}